\documentclass[aps, onecolumn, superscriptaddress]{revtex4-2}

\usepackage{amsmath,amssymb}
\usepackage{bm}
\usepackage{graphicx,epsfig}
\usepackage{lineno}
\usepackage[dvipsnames]{xcolor}

\usepackage[colorlinks=true,
            citecolor=blue,
            linkcolor=blue,
            urlcolor=blue]{hyperref}

\usepackage{ragged2e}
      
\begin{document}
\title{Longitudinal particle separation\color{black}}

\author{Siluvai Antony Selvan}
\email{antony.antonyravichandran@adelaide.edu.au}
\affiliation{School of Mathematical Sciences, Adelaide University, Adelaide, South Australia, Australia.}

\author{Rahil N Valani}
\affiliation{Rudolf Peierls Centre for Theoretical Physics, University of Oxford, OX1 3PU, United Kingdom.}

\author{Brendan Harding}
\affiliation{School of Mathematics and Statistics, Victoria University of Wellington, Wellington, New Zealand.}

\author{Yvonne M Stokes}
\affiliation{School of Mathematical Sciences, Adelaide University, Adelaide, South Australia, Australia.}

\date{\today}

\begin{abstract}
Owing to inertial effects, the flow through a three-dimensional curved duct focuses finite-sized spherical particles in the two-dimensional cross-section onto either stable equilibrium points or limit cycles. This hydrodynamic inertial focusing underpins various biomedical and industrial applications for size-based particle and cell sorting. Departing from conventional particle separation in the channel cross-section, we instead focus on particle separation in the primary flow direction, i.e., longitudinal separation. We consider
a duct with an elliptical centreline and a tall rectangular cross-section. For a given particle size, the nature of the cross-sectional equilibrium points depends on the local radius of curvature of the duct, and a periodical variation of the radius of curvature can result in a periodical bifurcation behaviour along its length. In particular, a duct geometry that undergoes a periodically varying saddle-node infinite-period (SNIPER) bifurcation can be used to improve longitudinal, at the expense of cross-sectional, particle clustering. For sufficiently large particles, this longitudinal clustering weakens at higher Reynolds numbers and with decreasing eccentricity, in contrast to small particles whose longitudinal clustering remains unaffected across a wide range of geometric configurations and flow conditions. Then, ducts with smaller eccentricities enable simultaneous separation along both the flow direction and the cross-section. In contrast, for larger eccentricities, the emergence of a SNIPER bifurcation promotes more pronounced longitudinal separation while compromising cross-sectional separation. These preliminary findings suggest that elliptically wound microfluidic devices might be used for longitudinal separation of particles by size, with potential biomedical and industrial applications.
\end{abstract}

\maketitle

\section{Introduction}\label{Intro}

Particles in Stokes flow remain confined to the streamlines of the flow due to the absence of inertia. In contrast, inertial migration is the movement across streamlines of particles in a flow under the influence of inertial lift forces \cite{harding2019effect,valani2022bifurcations}. In the case of flow within a duct, inertia drives particles towards stable attractors (i.e., points or closed curves) in the two-dimensional cross-section perpendicular to the primary direction of flow, and is often known as inertial particle focusing. Inertial migration was first observed experimentally by \citet{segre1961radial} in a straight pipe with circular cross-section; spherical particles of the same size that were initially uniformly distributed within the flow became focused to an annular region in the cross-section with a radius of 0.6 times the duct radius. This discovery has since paved the way for the field of inertial microfluidics, enabling transformative applications such as particle and cell sorting \cite{kuntaegowdanahalli2009inertial,lee2011inertial}, the isolation of circulating tumour cells (CTCs) \cite{sun2012double,hou2013isolation,warkiani2014slanted}, and flow cytometry \cite{bhagat2010inertial}. Despite these advances, most recent innovations in device design still rely predominantly on empirical prototyping \cite{di2009inertial,martel2014inertial,zhang2016fundamentals}, with new microfluidic geometries for specific applications fabricated and evaluated within days. Nevertheless, ongoing research is still required to predict and optimise particle sorting based on the underlying physical mechanisms.

Theoretical studies have been conducted to understand the focusing of non-deformable spherical particles in the cross-sections of two-dimensional and three-dimensional straight and curved ducts with varying cross-sectional geometries. For straight 2D channels and 3D circular pipes, the use of matched asymptotic expansions or the method of reflections provides an adequate description of particle migration \cite{ho1974inertial,schonberg1989inertial,matas2009lateral}. Unfortunately, these approaches become increasingly cumbersome, if not intractable, for three-dimensional ducts featuring curves and/or non-circular cross-sections. However, the application of a regular perturbation expansion and finite element methods has provided significant insights into the dynamics of lateral migration within straight/curved ducts with rectangular/trapezoidal cross-section, albeit at the expense of being applicable only for small to moderate Reynolds numbers. This framework has been successfully applied to particles suspended in straight ducts with rectangular cross-sections \cite{hood2015inertial,hood2016direct}. Building on this, \citet{harding2019effect} extended the methodology to ducts with constant bend radius and with rectangular and trapezoidal cross-sections at low flow rates. By adopting a frame of reference moving with the particle along the duct, the flow becomes steady while introducing inertial and, in the case of a curved duct, Coriolis forces into the Navier–Stokes equations. The bend radius was found to play a crucial role in determining the location of particle attractors within the cross-section. \citet{harding2019effect} further demonstrated that curved ducts with wide rectangular and trapezoidal cross-sections facilitate effective size-based particle separation. Subsequently, \citet{valani2024inertial} applied the same methodology to spiral ducts with rectangular cross-sections, in which the bend radius varies gradually along the primary flow direction, resulting in continuous variation of the particle attractors. Several other studies employ direct numerical simulations \cite{ookawara2006numerical,liu2016generalized} to model the flow fields in complex microchannels. Furthermore, \citet{ha2022dynamics} developed a reduced-order Zero Level Fit (ZeLF) model to investigate particle migration in curved ducts with rectangular cross-sections. This qualitative model successfully captures particle focusing behaviour in the cross-section of the duct and provides insight into how focusing and equilibrium bifurcations vary with flow and geometric parameters \cite{valani2022bifurcations}.

Cross-sectional particle focusing is a central principle in the design of inertial microfluidic devices, which are widely used for particle separation across diverse applications \cite{kuntaegowdanahalli2009inertial,sun2012double,bhagat2010inertial}. However, to the best of our knowledge, no microfluidic geometry has been reported that examines particle separation in the primary flow direction, i.e., along the duct. In the present study, we consider a duct featuring a centreline that is elliptical, but with small pitch such that it is an elliptical helix 
to prevent self-intersection after a full revolution. We also consider 
a uniform rectangular cross-section whose width in the direction of the bend radius is smaller than its height in the direction perpendicular to the bend radius. Similar to a spiral duct \cite{valani2024inertial}, the bend radius of an elliptical duct varies continuously; however, it does so in a periodic manner, distinguishing it fundamentally from the spiral geometry. Our analysis demonstrates that ducts with smaller eccentricities enable simultaneous longitudinal and cross-sectional separation. In contrast, ducts with larger eccentricities efficiently promote longitudinal separation but compromise cross-sectional separation.

The manuscript is organised as follows: Section \ref{MS} presents the model formulation, along with the first-order particle trajectory model that describes particle motion in the elliptical duct. The influence of geometry on particle equilibria and their dynamics is discussed in Section \ref{EG}. Sections \ref{PC} and \ref{PS} examine particle separation by size, both along the length of the duct and within the cross-section. Finally, Section \ref{Con} summarises the key findings of this study.

\section{Model setup}\label{MS}

\begin{figure}[htp]
  \centering
  \includegraphics[width=0.9\textwidth]{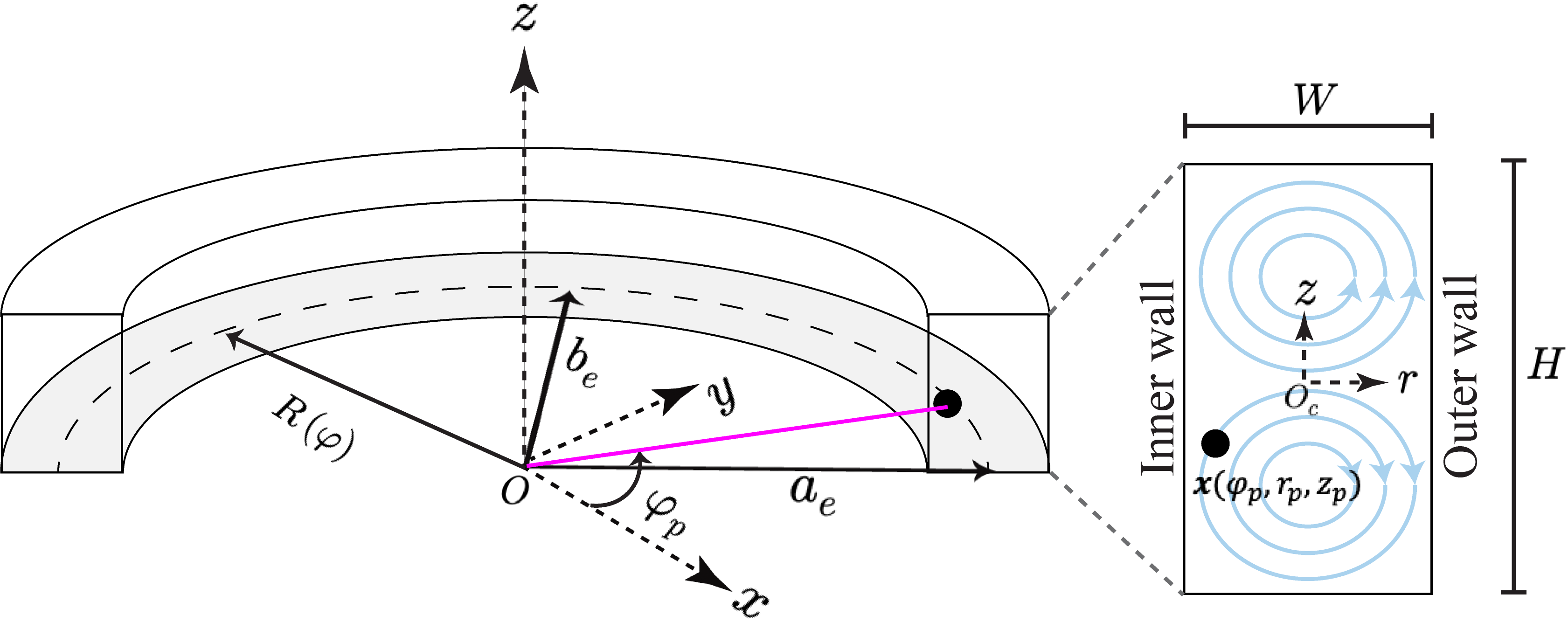}
  \caption{Model illustration of the particle of radius $a$ positioned at $\boldsymbol{x}_p=\boldsymbol{x}(\varphi_p,r_p,z_p)$ suspended in the fluid inside the duct with the elliptical centreline of radius $R(\varphi)$ (major axis $a_e$ and minor axis $b_e$) from origin $O$. The corresponding rectangular cross-section (shown in enlarged view) of width $W$ and height $H$, with the origin of local coordinates $(r,z)$ positioned at the duct's centre.}\label{f1}
\end{figure}

As shown in Fig.~\ref{f1}, a neutrally buoyant spherical particle of radius $a$ is suspended within a flow along a duct, whose centreline is considered to follow an ellipse with centre $O$ and major and minor radii $a_e$ and $b_e$, respectively.  The duct cross-section is rectangular with width $W$ and height $H$ such that the ratio $W:H=1:2$, and the cross-sectional coordinates in the horizontal and vertical directions are denoted by $r$ and $z$ with respect to the origin, $O_c$, positioned at the centre of the cross-section. Let $\varphi$ denote angular position around the duct from the $x$-axis, and let the centreline radius of the elliptical duct be given by
\begin{align}
    R(\varphi)=\frac{a_e\sqrt{1-e^2}}{\sqrt{\sin^2{\varphi}+(1-e^2)\cos^2{\varphi}}},\label{e2}
\end{align}
where $e:=\sqrt{1-b_e^2/a_e^2}$ is the eccentricity of the ellipse. The Cartesian coordinates of the particle (i.e., its centre) in the elliptical duct is expressed in terms of its angular $\varphi_p$, radial $r_p$, and vertical $z_p$ coordinates as,
\begin{align}
\boldsymbol{x}_p=\boldsymbol{x}(\varphi_p,r_p,z_p)=&\boldsymbol{x}_e(\varphi_p)+r_p\boldsymbol{n}(\varphi_p)+z_p\boldsymbol{e}_z,\label{e1}
\end{align}
where $\boldsymbol{x}_e(\varphi_p)=R(\varphi_p)(\cos{\varphi_p} \boldsymbol{e}_x+\sin{\varphi_p} \boldsymbol{e}_y)$ is the position of the duct centreline having normal vector \[\displaystyle \boldsymbol{n}(\varphi_p)=\varepsilon \cdot \frac{{d\boldsymbol{x}_e}/{d\varphi_p}}{{\|d\boldsymbol{x}_e}/{d\varphi_p}\|},
\] $\varepsilon$ being the 2D-Levi-Civita tensor \cite{astar2021new}. Here $\boldsymbol{e}_x$, $\boldsymbol{e}_y$ and $\boldsymbol{e}_z$ are the unit vectors in the $x$, $y$ and $z$ directions, respectively. 

In the absence of a particle, the velocity field of the flow through the duct, $\bar{\boldsymbol{u}}$, comprises the fully-developed axial component ($\bar{u}_{\varphi}$) together with a secondary flow ($\bar{u}_r$ and $\bar{u}_z$) which forms a pair of counter-rotating vortices in the duct cross-section induced by the curvature of the duct centreline \cite{dean1927note,dean1959note}. 
For analytical convenience, this flow is assumed to be driven by a pressure gradient in the main direction of flow. Of course, for a closed elliptical duct this is unphysical, but for the practical setting mentioned earlier, with a small pitch so that the duct is an elliptical helix, the flow may be driven by a pressure gradient whilst the pitch can be neglected as having negligible impact on the flow itself.
On introducing the particle, the background flow in the duct is disturbed, resulting in the disturbance velocity, $\boldsymbol{v}$, and pressure, $q$, (i.e., the difference between the fields in the presence and absence of the particle). To evaluate the background and disturbance fields, the elliptical duct is modelled as having a slowly varying radius of curvature $R_c(\varphi)$ along its length, in a similar spirit to our previous work on spiral ducts \cite{valani2024inertial}. Due to the elliptical geometry, we obtain periodic variations in $R_c$ which may be expressed as 
\begin{align}
    R_c(\varphi)=\frac{a_e}{\sqrt{1-e^2}} (\sin^2{\varphi}+(1-e^2) \cos^2{\varphi})^{3/2}.\label{e3}
\end{align}
It is important to emphasise that it is the radius of curvature which characterises the flow behaviour, not the radius (i.e. the distance from the origin) of the ellipse.
We solve the associated flow problems locally at each $\varphi_p$, where the local radius of curvature is compared to the circle with equivalent constant bend radius. For each value of $\varphi_p$, we obtain the cross-sectional forces ($F_r$ and $F_z$) and their corresponding drag ($C_r$ and $C_z$) acting on the suspended particle; a brief derivation of these quantities, following the work of \citet{harding2019effect}, is given in Appendix~\ref{App1}.

Using Stokes' law, we build the following first-order model for the particle trajectory by balancing both cross-sectional forces and viscous drag:
\begin{eqnarray}
    \frac{dr_p}{dt}=-Re_p\frac{F_r}{C_r},~~\frac{dz_p}{dt}=-Re_p\frac{F_z}{C_z},~~\mbox{and}~~\frac{d\varphi_p}{dt}=\frac{a\bar{u}_{\varphi_{p}}}{\sqrt{(R(\varphi_p))^2+(R'(\varphi_p))^2}}.\label{e4}
\end{eqnarray}
Here, the radial ($F_r$) and the vertical ($F_z$) cross-sectional forces depend on both the local radius of curvature $R_c(\varphi_p)$ and the particle radius $a$. The non-dimensional parameter $Re_p=Re(a/l)^2$ is known as the particle Reynolds number, where $Re=\rho U_m l/\mu$ is the flow Reynolds number corresponding to the fluid of density $\rho$ and viscosity $\mu$, with $l=\min\{W, H\}= W$ and $U_m$ denoting the reference length and velocity scales, respectively. A detailed derivation for obtaining the axial velocity component of the first-order model (i.e., $d\varphi_p/dt$) is given in Appendix~\ref{App2}. Equation~\eqref{e4} is solved using the ode15s solver (relative tolerance$~=10^{-3}$ and absolute tolerance$~=10^{-6}$) in MATLAB, with particles initially released in the cross section at $\varphi=0^\circ$. Within this cross-section, we employ the MATLAB random number generator (initialised with unit seed, i.e. \texttt{rng(1)}) to randomly position the particles. Importantly, we verified that the reported results remain qualitatively unchanged for other choices of the seed.

In this model, we neglect particle-particle interactions, implying that the results are valid in the regime of low particle density. Furthermore, we evaluate the particle equilibrium and its stability within the cross-section. To obtain the particle equilibrium, we look for the positions $(r^*,z^*)$, where the cross-sectional forces vanish (i.e., $F_r=F_z=0$). The corresponding stability of these equilibrium positions $(r^*,z^*)$ is determined from the following Jacobian matrix:
\begin{align}
    J=\begin{pmatrix}
\partial F_r/\partial r & \partial F_r/\partial z \\
\partial F_z/\partial r & \partial F_z/\partial z
\end{pmatrix},\label{e5}
\end{align}
where the nature of the eigenvalues ($\lambda_1=\lambda_{1r}+\mathrm{i}\lambda_{1i}$ and $\lambda_2=\lambda_{2r}+\mathrm{i}\lambda_{2i}$) determines the type of particle equilibrium. The periodic variation of the duct’s radius of curvature along the elliptical path leads to periodic variations in the number, nature, and location of equilibrium points for suspended particles. 
\section{Particle behaviour in the cross-section}
\label{EG}
We begin our investigation by examining the effect of an elliptical duct centreline on particle equilibria and their motion in the rectangular cross-section. The results presented below are non-dimensionalised using half the width of the rectangular cross-section as follows: $\tilde{a}=2a/W$, $\tilde{a}_e=2a_e/W$, $\tilde b_e=2b_e/W$, $\tilde{r}=2r/W$, $\tilde{z}=2z/W$, $\tilde{R}=2R/W$, and $\tilde{R}_c=2R_c/W$. 

\subsection{Particle equilibria}\label{PE}

\begin{figure}[htp!]
  \centering
  \includegraphics[width=0.9\textwidth]{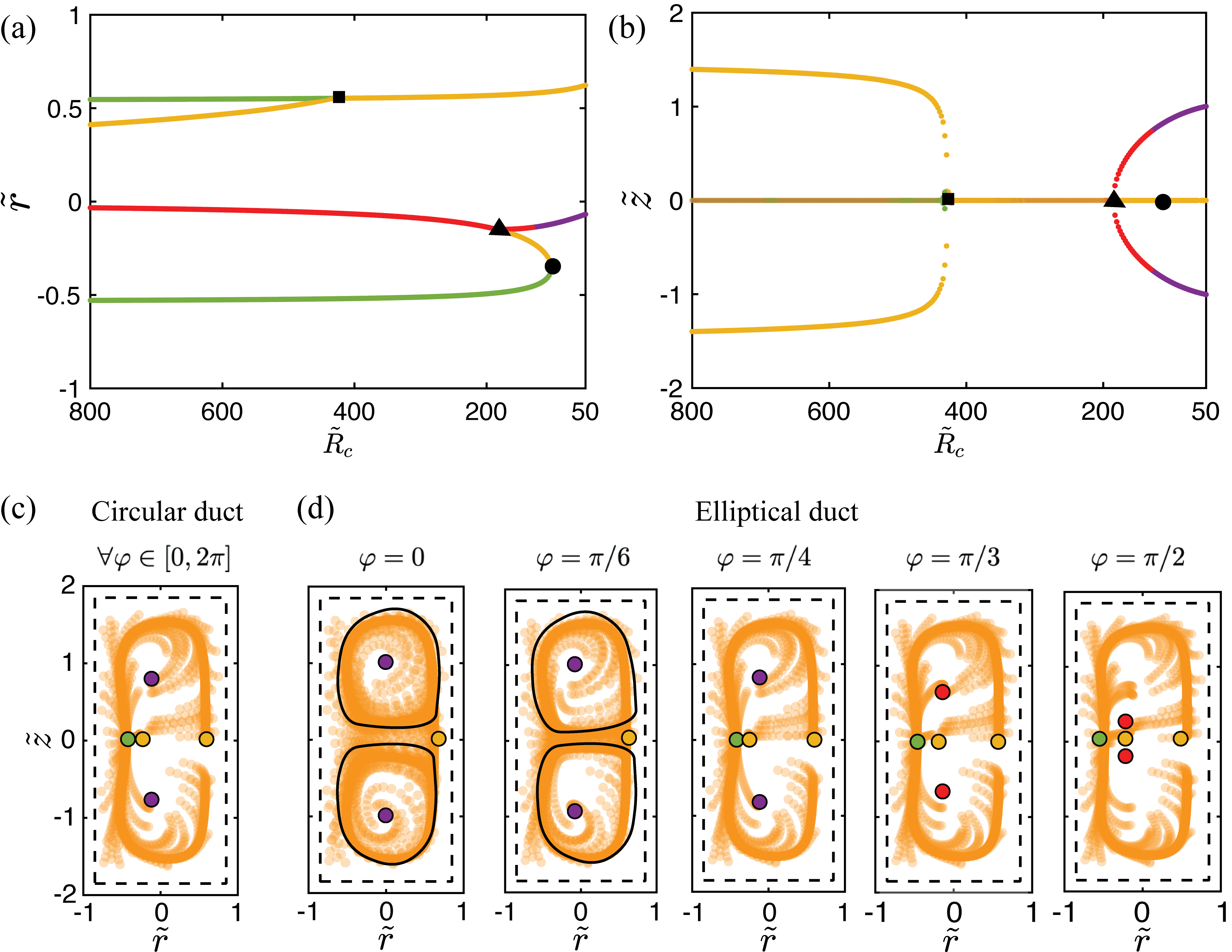}
  \caption{(a,b) 
  Bifurcation diagrams showing the (a) horizontal $\tilde r$ and (b) vertical $\tilde z$ coordinates of equilibria as a function of the radius of curvature $\tilde R_c$, which decreases from left to right. The particle is of radius $\tilde{a}=0.15$ and the duct has a $1\times2$ rectangular cross-section. The stable nodes, saddle points, unstable nodes, and unstable spiral points are shown in green, yellow, red, and purple, respectively. The black square, triangle, and circle denote the locations of subcritical pitchfork, supercritical pitchfork, and saddle-node infinite-period (SNIPER) bifurcations, respectively. (c,d) Particle equilibria in the cross-section of (c) circular ($\tilde{a}_e=120$, $e=0$, $\tilde R_c=120$) and (d) elliptical ($\tilde{a}_e=120$, $e=0.75$, $52.5\le\tilde R_c\lesssim 181.423$) ducts. As in (a,b), the different colours denote stable nodes, saddle points, unstable nodes, and unstable spiral points, but with black outlines to clearly differentiate them from the example particle trajectories, which are shown in orange for $Re=100$. Note that trajectories were computed assuming the motion at any $\varphi$ is locally approximated by that in a constant-curvature duct.
  Limit cycles are shown as black solid curves in panel (d). The dashed black lines denote locations of particle centres having surfaces that touch the wall of the duct (solid black lines).}\label{f2}
\end{figure}

We start by classifying the particle equilibria as a function of the radius of curvature of the duct, $\tilde R_c$. We consider the range $\tilde{R}_c \in [50,800]$ to demonstrate the various bifurcations for a particle of radius $\tilde{a}=0.15$. The bifurcation diagrams of
Fig.~\ref{f2}(a,b) show the $\tilde r$ and $\tilde z$ coordinates of the equilibria versus $\tilde R_c$, respectively. 

In these diagrams, stable nodes, saddle points, unstable nodes, and unstable spirals are indicated by the colours green, yellow, red, and purple, respectively. Note that $\tilde R_c$ decreases from left to right. The particle behaviour, along with the nature of the eigenvalues at these equilibrium points, is given in Table~\ref {T1}. 
\bgroup
\def\arraystretch{1.5}%
\begin{table}[htp!]
\begin{footnotesize}
    \centering
    \begin{tabular}{|c|c|c|c|}
    \hline
        \textbf{Eigenvalues ($\lambda$)} & \textbf{Equilibrium type} & \textbf{Particle behaviour} \\ 
        \hline
        $\lambda_1, \lambda_2 \in \mathbb{R}$, both negative & Stable node & Particles converge to the equilibrium position.\\ \hline
         $\lambda_1, \lambda_2 \in \mathbb{R}$, both positive & Unstable node & Particles diverge away from the equilibrium position.\\ \hline  $\lambda_1, \lambda_2 \in \mathbb{R}$, 
          opposite signs & Saddle point & Stable in one direction, 
          unstable in the other
         direction.\\ \hline
         $\lambda_1, \lambda_2 \in \mathbb{C}$, negative real part & Stable spiral & Particles spiral into the equilibrium position.\\ \hline
         $\lambda_1, \lambda_2 \in \mathbb{C}$, positive real part & Unstable spiral & Particles spiral away from equilibrium position.\\ \hline
    \end{tabular}
\caption{\justifying{Type of particle equilibrium and the corresponding particle behaviour at the equilibrium position $(r^*,z^*)$ as determined by
the eigenvalues ($\lambda_1$ and $\lambda_2$).}}\label{T1}
\end{footnotesize}
\end{table}

As the radius of curvature $(\tilde{R}_c)$ decreases, a subcritical pitchfork bifurcation is encountered (marked by a black square). Here, the vertically symmetric saddle points near $\tilde r=0.5$, which align vertically with the stable node at $\tilde z=0$, merge with it to form a single saddle point located at $\tilde z=0$ and which remains close to the outer wall ($\tilde r\approx 0.5$, Fig.~\ref{f2}(a,b)). With further decrease in $\tilde{R}_c$ (right of the black square), the unstable node near the cross-sectional centre $(\tilde r,\tilde z)=(0,0)$ drifts slightly towards the inner wall. Then, it undergoes a supercritical pitchfork bifurcation (marked by black triangles in Fig.~\ref{f2}(a,b)), where the unstable node changes into a new saddle point along with two unstable nodes. As $\tilde R_c$ decreases further, these two unstable nodes diverge vertically in the cross-section and become unstable spirals, as shown in Fig.~\ref{f2}(b). On the other hand, the newly created saddle point persists for a time on the vertical centreline, moving towards the inner duct wall with decreasing $\tilde R_c$. However, this does not persist indefinitely. As $\tilde{R}_c$ decreases further, it collides with the stable node on $\tilde z=0$ near the inner wall and vanishes in a saddle-node infinite-period (or SNIPER) bifurcation  
(marked by black circles.) This sequence of bifurcation events\textemdash subcritical pitchfork, followed by supercritical pitchfork, and ultimately a SNIPER bifurcation\textemdash highlights the complex restructuring of equilibria induced by varying the radius of curvature \cite{valani2022bifurcations}. For a smaller particle radius, there is a similar bifurcation diagram to that shown in Fig.~\ref{f2}, but it shifts towards the left, i.e., events occur at larger radii of curvature. For example, for the particle of radius $\tilde{a}=0.05$, the onset of subcritical pitchfork (square marker), supercritical pitchfork (triangular marker), and SNIPER (circular marker) bifurcation is as seen in
Fig.~\ref{sf1}(a,b) of Appendix~\ref{App3}. 

We now examine how the radius of curvature of the duct centerline governs the particle motion in the cross-section for a flow with Reynolds number $Re=100$. We consider a circular ($\tilde{a}_e=120$ and $e=0$, $\tilde R_c=120$) and an elliptical ($\tilde{a}_e=120$ and $e=0.75$, $52.5\le\tilde R_c\lesssim 181.423$) duct; equilibria and particle trajectories are
shown in Fig.~\ref{f2}(c) and (d), respectively. In a circular duct, due to the constant radius of curvature (i.e., $\tilde{R}_c=\tilde{a}_e,~\forall \varphi$), the positions and nature of the particle equilibria don't change with the orbital angle $\varphi$. For example, as seen in Fig.~\ref{f2}(c), for the circular duct of radius $\tilde{a}_e=120$, we observe a stable node, saddle points, and unstable spirals, whose location and nature remain the same for all $\varphi\in [0,~2\pi]$, and particle trajectories evolve according to these. On the other hand, the elliptical duct has a periodically varying radius of curvature given by Eq.~\eqref{e3},
which varies within the range 
\begin{align}
    \tilde{R}_c\in \bigg[(1-e^2)\tilde{a}_e,~\frac{\tilde{a}_e}{\sqrt{1-e^2}}\bigg]. 
\end{align}
 
As $\tilde{R}_c$ oscillates between its minimum and maximum values, the equilibrium positions of a particle vary accordingly. Figure~\ref{f2}(d) illustrates this behaviour for $\tilde{a}_e=120$ and $e=0.75$, where the particle trajectories at each $\varphi$ are assumed to evolve according
to a locally approximated constant-curvature duct. At $\varphi=0$, where the radius of curvature is at its minimum value, the cross-section contains unstable spirals surrounded by stable limit cycles coexisting with a saddle point. As $\varphi$ increases to $\pi/2$ and the radius of curvature grows, the system evolves in a manner analogous to moving leftward in the bifurcation diagrams given in Fig.~\ref{f2}(a,b) (up to $\tilde{R}_c\approx 182 $). Consequently, as $\varphi$ increases from $0$ through $\pi/6$ towards $\pi/4$, the location of the unstable spiral changes while the surrounding limit cycle grows a little in size. Near to $\varphi=\pi/4$ there is a bifurcation with the birth of a stable node and saddle point pair on the centreline $\tilde z=0$, towards the inner duct wall, via a SNIPER bifurcation. As $\varphi$ increases even more, this pair separates along $\tilde z=0$, and the unstable spirals become unstable nodes (as seen in the panel for $\varphi=\pi/3$), which then move close to the saddle point on $\tilde z=0$ near the cross-sectional centre (see panel for $\varphi=\pi/2$). For $\pi/2\le\phi\le\pi$, the radius of curvature decreases, and the equilibria change in the reverse order, and so on. This process of qualitative change in particle equilibria along the elliptical duct reflects the strong interplay between geometric modulation and dynamical stability.

\begin{figure}[t!]
  \centering
  \includegraphics[width=0.9\textwidth]{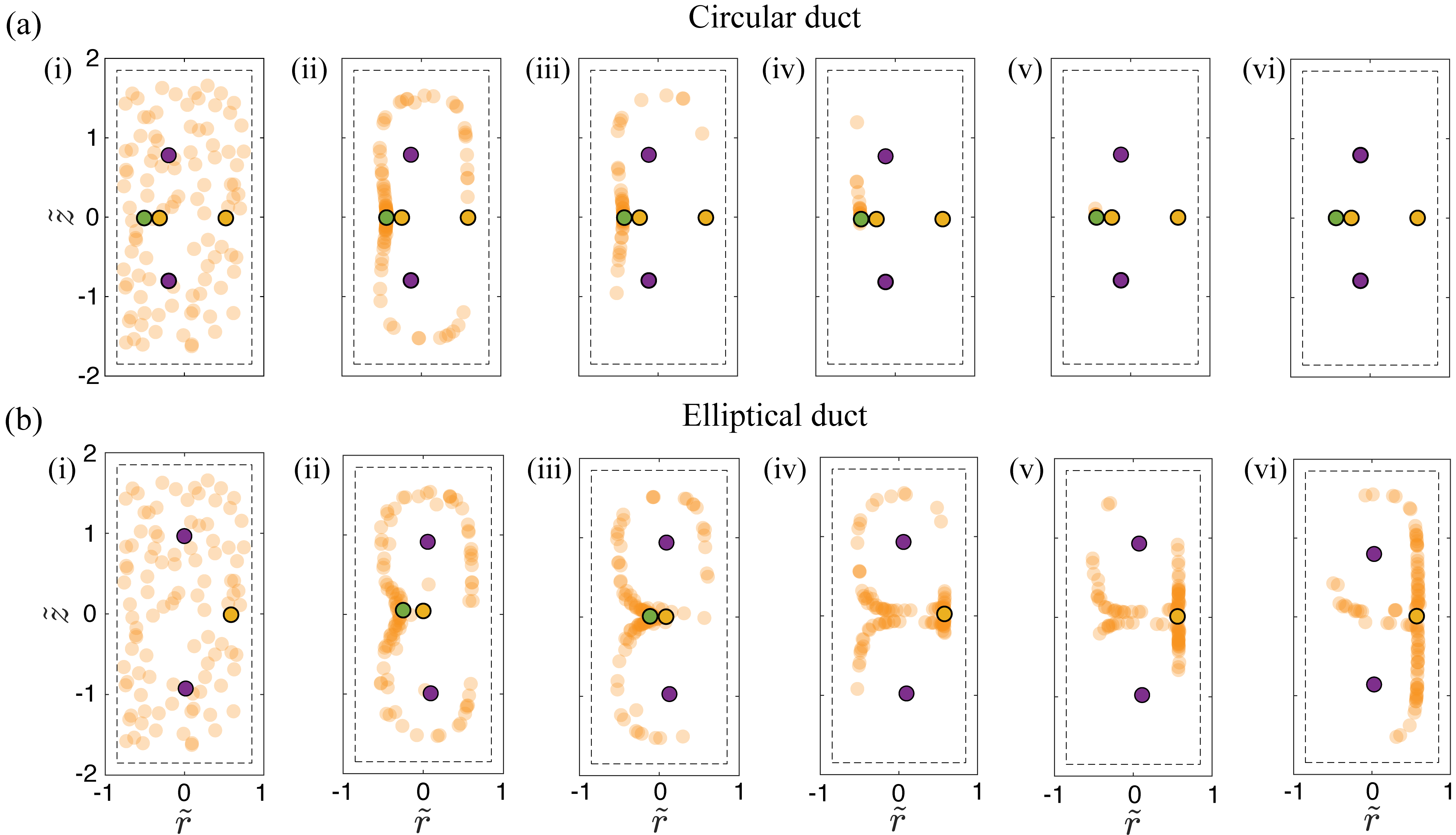}
  \caption{Time-dependent motion in the cross-section of $100$ initially randomly distributed particles of radius $\tilde a=0.15$ (orange circles) 
  along with their equilibria for $Re=100$. (a)  A circular duct ($\tilde{a}_e=120$, $e=0$) at times (i) $\tilde{t}=0$, (ii) $\tilde{t}=50$, (iii) $\tilde{t}=100$, (iv) $\tilde{t}=150$, (v) $\tilde{t}=200$, and (vi) $\tilde{t}=250$. (b) An elliptical duct ($\tilde{a}_e=120$, $e=0.75$) at times (i) $\tilde{t}=0$, (ii) $\tilde{t}=20$, (iii) $\tilde{t}=40$, (iv) $\tilde{t}=60$, (v) $\tilde{t}=80$ and (vi) $\tilde{t}=100$. The stable node, saddle points, and unstable spiral points are represented by green, yellow, and purple circles, but with black outlines to clearly differentiate them from the particle positions.  The dashed black lines denote locations of particle centres having surfaces that touch the wall of the duct (solid black lines).} \label{f3}
\end{figure}

\subsection{Particle dynamics in the cross-section}\label{PT}

We now explore the effect of the radius of curvature on the time-dependent motion of particles ($\tilde{a}=0.15$) in the rectangular cross-section. Again, we use a flow Reynolds number of $Re=100$. As outlined in Section~\ref{MS}, the particles are released in both circular and elliptical ducts at $\varphi_p=0$, with their initial positions randomly distributed within the cross-section (Fig.~\ref{f3}(a, i) and (b, i)). 

Tracking the particles in the circular duct cross-section, Fig.~\ref{f3}(a), we notice the following dynamical pattern: particles are initially pushed away from the cross-sectional centre due to the inertial lift force (Fig.~\ref{f3}(a, ii)) and drawn into spiral trajectories governed by the unstable spiral, which guide them toward the stable node located near the inner wall (Fig.~\ref{f3}(a, iii-vi)). This migration process is relatively slow, requiring a significant amount of time ($\tilde{t}=250$) for all particles to fully converge to the stable node. 

In contrast, the motion of particles in the elliptical duct, Fig.~\ref{f3}(b), exhibits different dynamics. Initially, the behaviour resembles that of the circular duct (Fig.~\ref{f3}(a, ii) and (b, ii)), with particles being repelled from the cross-sectional centre. Note that, in Fig. 3(b), because particles do not all lie within a single cross-section, we have shown particle equilibria for a cross-section representative of that pertaining to the cross-sections occupied by most of the particles. In general, particles moving along the length of the elliptical duct 
encounter periodically varying equilibria. At $\tilde{t}=20$, 
particles located near the centre of the cross-section, and to the left of the saddle node, are attracted toward the stable node close to the inner wall (Fig.~\ref{f3}(b, i–ii)). As the stable node migrates toward the saddle point, these particles move with it towards the cross-sectional centre (Fig.~\ref{f3}(b, iii)). When these equilibria collide and vanish in a SNIPER bifurcation, the particles that were held near the inner wall move across the vertical centre of the duct towards the outer wall.
However, the saddle near the centre of the outer wall redirects the particles into the stable limit cycles, as shown in Fig.~\ref{f3}(b, iv–vi). This SNIPER bifurcation occurs periodically along the elliptical length, with the stable node and central saddle vanishing and reappearing. In the present study, we are interested in understanding the particle motion in the longitudinal flow direction in response to this. In the following section, we analyse this behaviour in detail and systematically examine the influence of geometric and flow parameters. 

\section{Longitudinal particle motion} \label{PC}

Building on the previous Section~\ref{PT}, we now investigate how flow conditions and the geometry of the duct centreline affect particle motion longitudinally, i.e., in the direction of the primary flow along the duct. We are particularly interested in the influence of the SNIPER bifurcation. We will show results for large ($\tilde a=0.15$) and small ($\tilde a=0.05$) particles, which will be distinguished in the figures by different marker sizes and colours. 

 \begin{figure}[!t]
  \centering
  \includegraphics[width=0.94\textwidth]{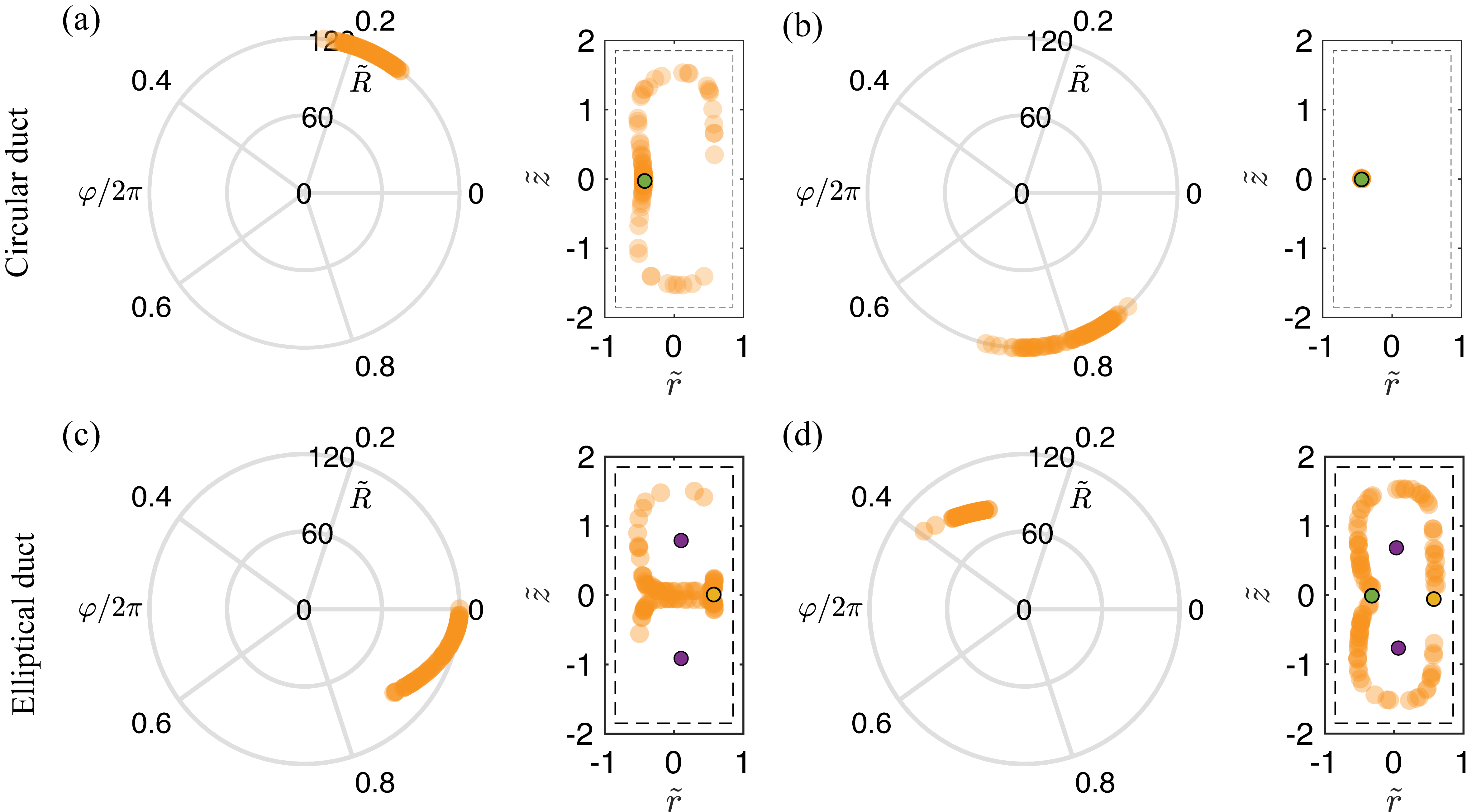}
  \caption{Time-dependent motion of $100$ initially randomly distributed 
  particles of radius $\tilde{a}=0.15$ (orange circles) suspended in a flow with $Re=150$ shown for (a,b) a circular ($\tilde{a}_e=120$, $e=0$) and (c,d) an  elliptical ($\tilde{a}_e=120$, $e=0.75$) duct. The particle positions are shown in both the polar $(\tilde{R},\varphi)$ and cross-sectional $(\tilde{r},\tilde{z})$ planes at times (a) $\tilde{t}=100$ and (b) $\tilde{t}=300$ for the circular duct, and (c) $\tilde{t}=420$ and (d) $\tilde{t}=600$ for the elliptical duct. The stable node, saddle points, and unstable spirals are represented by green, yellow, and purple circles with black outlines, respectively. The dashed black lines in the cross-sectional plots denote locations of particle centres having surfaces that touch the wall of the duct (solid black lines).  
  Simulations showing positions of the particles over the time interval, $\tilde{t}\in [0, 1500]$ are given in Supplementary Movies~\ref{SM1} and~\ref{SM2} for the circular and elliptical ducts, respectively.}\label{f4}
\end{figure}

\subsection{Effect of centreline ellipticity}

We begin by examining the effect of the ellipticity of the duct centreline (neglecting the small pitch) on the longitudinal particle motion, as shown in Fig.~\ref{f4}. Both circular ($e=0$, Fig.~\ref{f4}(a)) and elliptical ($e=0.75$, Fig.~\ref{f4}(b)) ducts of radius $\tilde{a}_e=120$, having the large particles suspended in a flow of high Reynolds number ($Re=150$), are shown. For both ducts, the particles quickly focus to a slow manifold in the cross-section as the longitudinal length of the particles increases. In the case of the circular duct (Fig.~\ref{f4}(a)), examination of the cross-sectional dynamics then reveals that the particles converge towards a stable equilibrium near the inner wall. During this evolution, the longitudinal 
length of the particle distribution further increases. Once the focusing to the cross-sectional stable node is complete, the length of the particle distribution along the duct remains constant (see Supplementary Movie~\ref{SM1}). In the case of the elliptical duct (Fig.~\ref{f4}(c)), 
the longitudinal particle distribution widens when the stable node vanishes (in a SNIPER bifurcation with the approximately central saddle node) and the particles move quickly across the cross-sectional centre $\tilde z=0$ towards the saddle point near the centre of the outer duct wall. However, the particles then aggregate in the longitudinal direction as they 
become vertically distributed close to the outer duct wall 
and move along the stable orbits around the unstable spiral points back towards the stable focus, which reappears (Fig.~\ref{f4}(d); see also Supplementary Movie~\ref{SM2}). 

\begin{figure}[!t]
  \centering
  \includegraphics[width=0.94\textwidth]{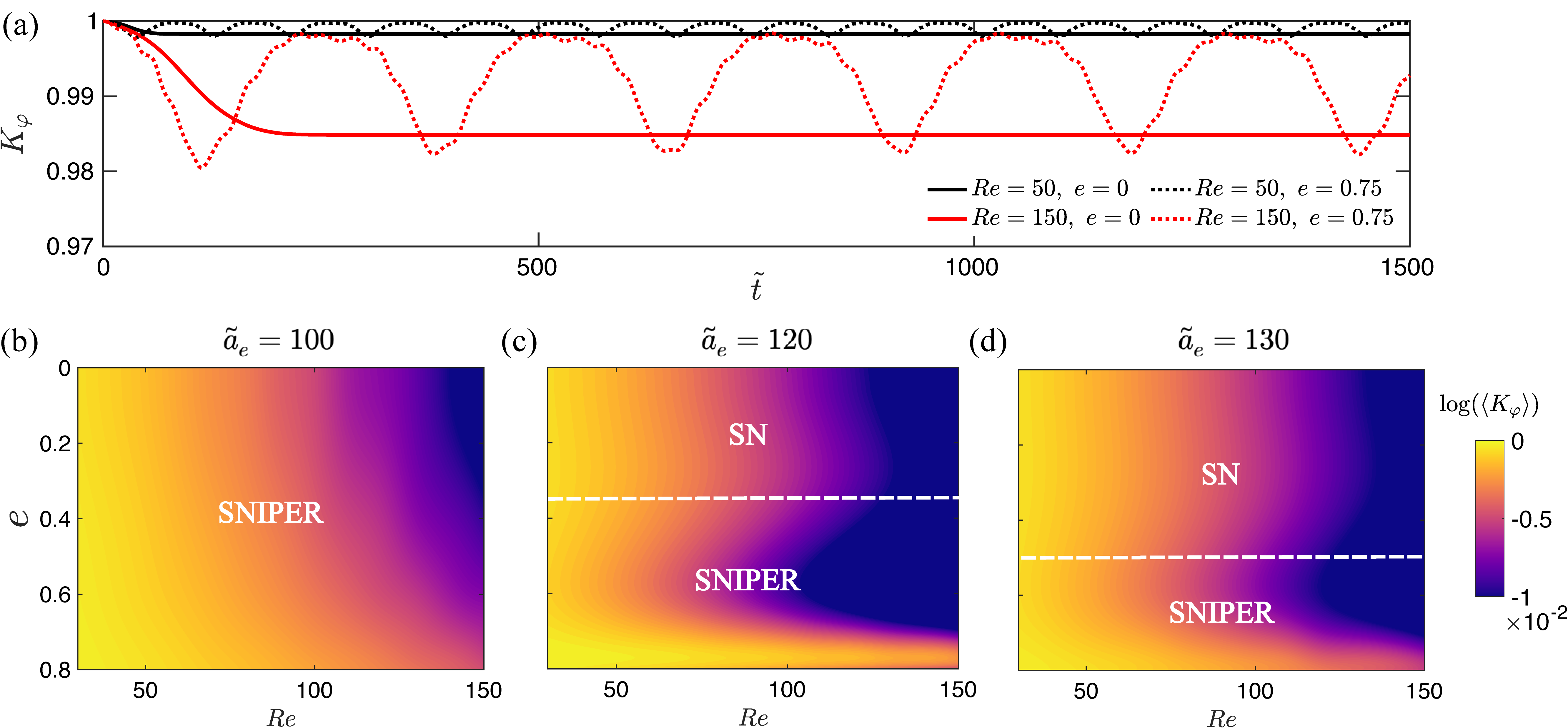}
  \caption{(a) Evolution in time $\tilde t$ of the Kuramoto order parameter $K_{\varphi}$, evaluated over $N=100$ particles of size $\tilde{a}=0.15$, for the circular ($\tilde{a}_e=120$, $e=0$, solid linestyle) and elliptical ($\tilde{a}_e=120$, $e=0.75$, dotted linestyle) 
  ducts, showing the effect of the flow Reynolds number $Re$. 
  (b--d) Contour maps corresponding to the log of time-averaged Kuramoto order parameter, $\log(\langle K_{\varphi}\rangle)$ over $\tilde{t}=1500$, as a function of $Re$ and eccentricity $e$, for various major axis radii: (b) $\tilde{a}_e=100$, (c) $\tilde{a}_e=120$, and (d) $\tilde{a}_e=130$. For each $(e,~Re)$ combination, $\log(\langle K_\varphi\rangle)$ is evaluated for $N=100$ particles of size $\tilde a_e$, initially released at random positions in the cross-section $\varphi=0$.  SN denotes the region in which there is a persistent stable node and saddle point combination near the inner duct wall;  SNIPER denotes the region in which this stable node and saddle point combination undergoes infinite-period bifurcations; the white-dashed line marks the division between these two regions. The colour bar shows the range of $\log(\langle K_{\varphi}\rangle)$.}\label{f5}
\end{figure}

To quantify such longitudinal clustering, we introduce an order parameter known as the Kuramoto order parameter \cite{dorfler2011critical}{,}
\begin{align}
    K_{\varphi}(\tilde t)=\bigg|\frac{1}{N}\sum_{p=1}^{N}\exp({\mathrm{i}\varphi_p(\tilde{t})})\bigg|, \label{e7}
\end{align}
where $\mathrm{i}=\sqrt{-1}$,~$\varphi_p(\tilde t)$ is the orbital position of the $p$-th particle at time $\tilde t$, and $N$ is the number of particles in the duct. The parameter satisfies $K_{\varphi}\in [0, 1]$, with values approaching zero indicating weak longitudinal aggregation and values approaching unity denoting strong longitudinal aggregation.  Fig.~\ref{f5}(a) shows the time evolution of the Kuramoto order parameter, $K_{\varphi}$, for $N=100$ large particles ($\tilde{a}=0.15$) for both circular and elliptical ducts at two different values of the Reynolds number. Results for the circular duct are denoted by solid lines, while those for the elliptical duct are shown by dotted lines. All particles are initially released at $\varphi=0$, resulting in values of $K_{\varphi}$ close to unity. For the circular duct, the order parameter decreases over early time $\tilde t$, but then attains a constant value which is smaller for larger $Re$. Thus, there is weaker particle clustering with enhanced inertial effects (larger $Re$). In contrast, for the elliptical duct (dotted lines), the order parameter initially decreases and subsequently exhibits oscillatory behaviour with increasing $\tilde{t}$. Thus, there are times of greater and lesser particle clustering, with the magnitude of the oscillation increasing with $Re$.

In Fig.~\ref{f5}(b--d), the log of the time-averaged Kuramoto order parameter, $\log(\langle K_{\varphi} \rangle)$, is shown in  the duct ellipticity ($e$) versus flow Reynolds number ($Re$) plane for varying major axis radii, $\tilde{a}_e$. Across all geometries, $\log(\langle K_{\varphi} \rangle)$ systematically decreases with fixed $e$ and increasing $Re$, indicating a progressive weakening of longitudinal particle aggregation as inertial effects intensify. For $\tilde{a}_e=100$ (Fig.~\ref{f5}(b)), there are SNIPER bifurcations for all parameter sets shown in the $(e, Re)$ space, i.e., the periodic appearance and disappearance of a stable node and saddle point near the inner duct wall. Moreover, increasing the eccentricity, $e$, enhances longitudinal particle aggregation due to the increased range in the radius of curvature, which results in particles remaining near the outer wall of the cross-section for extended durations (see Supplementary Movie~\ref{SM2}). 

However, for sufficiently large major-axis radius $\tilde a_e$, there are two distinct dynamical regimes:  a stable-node-dominated regime denoted SN, where the duct centreline is sufficiently close to circular ($e$ sufficiently small), and a SNIPER-dominated regime for larger $e$, Fig.~\ref{f5}(c,d). Considering first the case $\tilde a_e=120$ (Fig.~\ref{f5}(c)), in the stable-node regime $e<0.35$, particle aggregation dynamics closely resemble those shown in Fig.~\ref{f4}(a,b) for a circular duct, both in the longitudinal direction and in the cross-section, with the Reynolds number serving as the primary control parameter. In contrast, within the SNIPER regime $e\geq0.35$, $\log(\langle K_{\varphi} \rangle)$ attains its maximum values at larger eccentricities or low Reynolds numbers. Notably, for $e\gtrsim 0.75$, $\langle K_{\varphi} \rangle$ remains close to unity (i.e., $\log(\langle K_{\varphi} \rangle) \to 0$) even at comparatively higher $Re$ 
compared with the stable-node regime. Further increasing the major-axis radius to $\tilde{a}_e=130$ shifts the boundary of the SNIPER regime (white-dashed line) toward higher eccentricities, thereby reducing the parameter space within the displayed plot domain over which SNIPER bifurcations occur. Taken together, these observations demonstrate that SNIPER bifurcations play a central role in promoting and sustaining strong longitudinal particle clustering in the duct. 

In practice, particles will be released within a finite longitudinal region rather than at a single longitudinal position $\varphi=0$. The effect of this is considered in Appendix~\ref{App4}.  A similar particle-clustering pattern is observed for different axial initial conditions (see Appendix~\ref{App4}), with the maxima of oscillation in $K_{\varphi}(\tilde{t})$ equal to its initial value, $K_{\varphi}(0)$.

\begin{figure}[h!]
  \centering
  \includegraphics[width=0.9\textwidth]{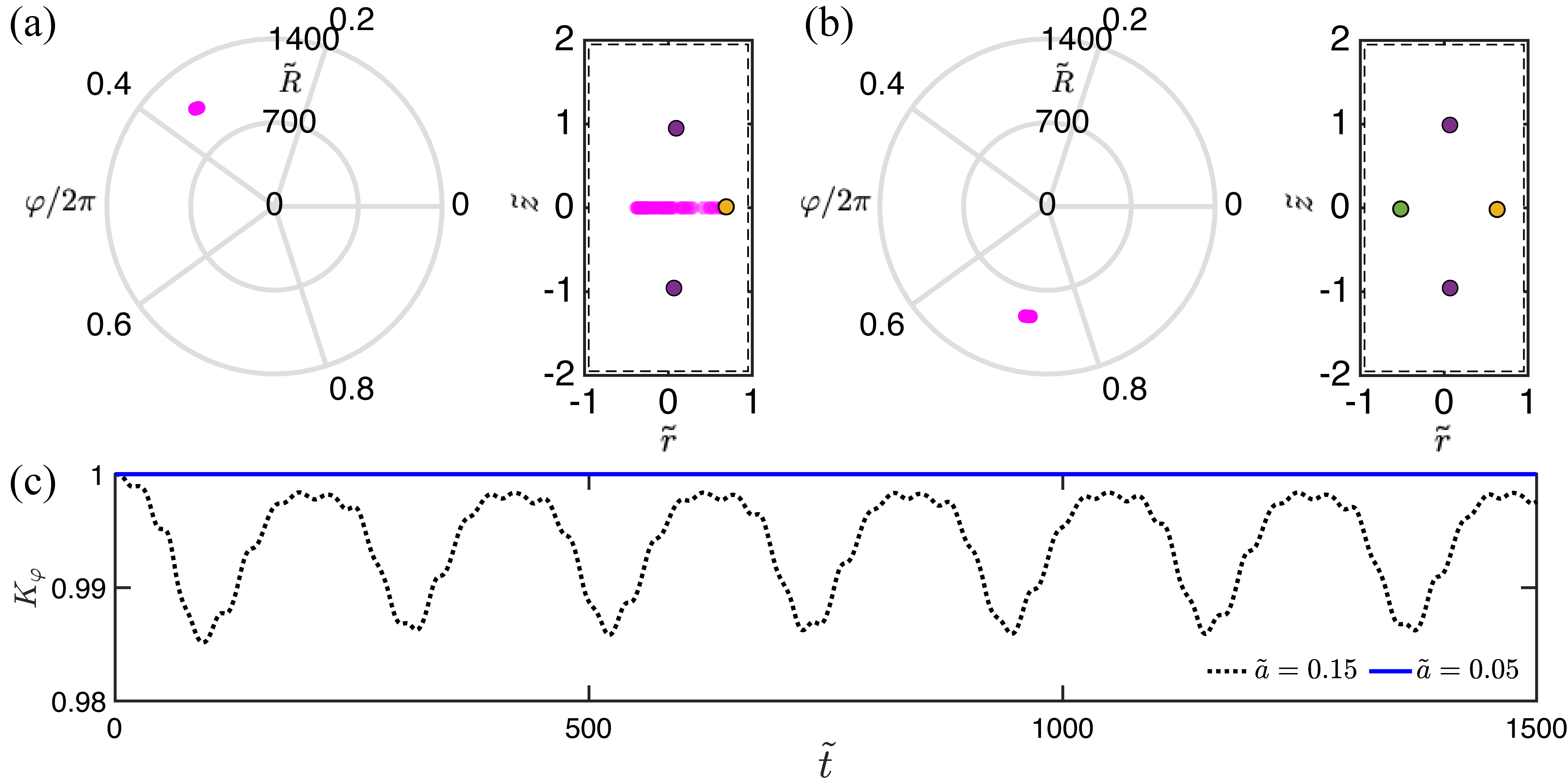}
  \caption{(a,b) Time-dependent motion of $100$ small particles, initially randomly distributed in the cross-section $\varphi=0$, of radius $\tilde{a}=0.05$ (magenta circles) suspended in a flow with $Re=150$ in an elliptical ($\tilde{a}_e=1400$ and $e=0.75$) duct. Particle distribution at (a) $\tilde{t}=1810$ and (b) $\tilde{t}=4000$ (all particles under green circle in cross-section), shown in both the polar $(\tilde{R},\varphi)$ and cross-sectional $(\tilde{r},\tilde{z})$ planes. The stable node, saddle points, and unstable spiral points are represented by green, yellow, and purple circles with black outlines, respectively. The dashed black lines in the cross-sectional plots denote locations of particle centres having surfaces that touch the wall of the duct (black solid lines).  The black arrows show the direction of particle motion in the cross-section. The complete simulation over the time interval $\tilde{t}\in [0, 15000]$ is given in Supplementary Movie~\ref{SM3}. (c) Evolution in time $\tilde t$ of the Kuramoto order parameter, $K_{\varphi}$, evaluated over $N=100$ particles, for large particles $\tilde{a}=0.15$ with $\tilde{a}_e=120$ (also shown in Fig.~\ref{f5}(a)), and for small particles $\tilde{a}=0.05$ with $\tilde{a}_e=1400$. Both ducts have an eccentricity of $e=0.75$.}\label{f6}
\end{figure}

\subsection{Effect of particle size}

We next examine the effect of particle size on the longitudinal clustering. Specifically, we consider smaller particles than those considered so far, having a radius $\tilde{a}=0.05$, for which the SNIPER bifurcation occurs at a relatively large radius of curvature $\tilde{R}_c=1550$, Fig.~\ref{sf1}(a,b). We calculate the time-dependent motion of $100$ of these particles, initially randomly suspended in the cross-section at $\varphi=0$, in a duct having major-axis radius $\tilde{a}_e=1400$ and ellipticity $e=0.75$, with a flow having Reynolds number $Re=150$. The particle locations at times $\tilde{t}=1810$ and $\tilde{t}=4000$ are shown in Fig.~\ref{f6}(a,b), respectively; note that all particles are under the green stable node in the cross-sectional plot of Fig.~\ref{f6}(b). The complete spatio-temporal evolution of the particle positions in both the polar $(\tilde{R},\varphi)$ and cross-sectional $(\tilde{r},\tilde{z})$ planes is provided in the Supplementary Movie~\ref{SM3}. At $\tilde{t}=1810$ (Fig.~\ref{f6}(a)), many particles 
are distributed across the centre of the cross-section $\tilde z=0$, due to the very recent disappearance of a stable node near the inner wall (i.e., a SNIPER bifurcation). As a result, particle aggregation in the $(\tilde{R},\varphi)$ plane is expected to weaken; however, such an effect is negligible as compared with the significant loss of clustering observed in the case of a large particle (Fig.~\ref{f4}(c)). In fact, there is little difference in the longitudinal clustering in Fig.~\ref{f6}(a,b), despite the different cross-sectional distribution of particles. 
On quantifying the particle clustering in the longitudinal direction (Fig.~\ref{f6}(c)), we notice that the value of $K_{\varphi}$ for the small particles remains close to unity for all $\tilde{t}$, as opposed to the large particles. 

This is true even when we examine the time-averaged Kuramoto order parameter (Fig.~\ref{sf1}(c,d)) in the $(e, Re)$ plane for various major axis radii, $\tilde{a}_e$.  Thus, the longitudinal clustering for small particles remains unaffected by the changing bend radius, indicating that large and small particles might be longitudinally separated.

\section{Particle separation} \label{PS}

\begin{figure}[!b]
  \centering
  \includegraphics[width=0.9\textwidth]{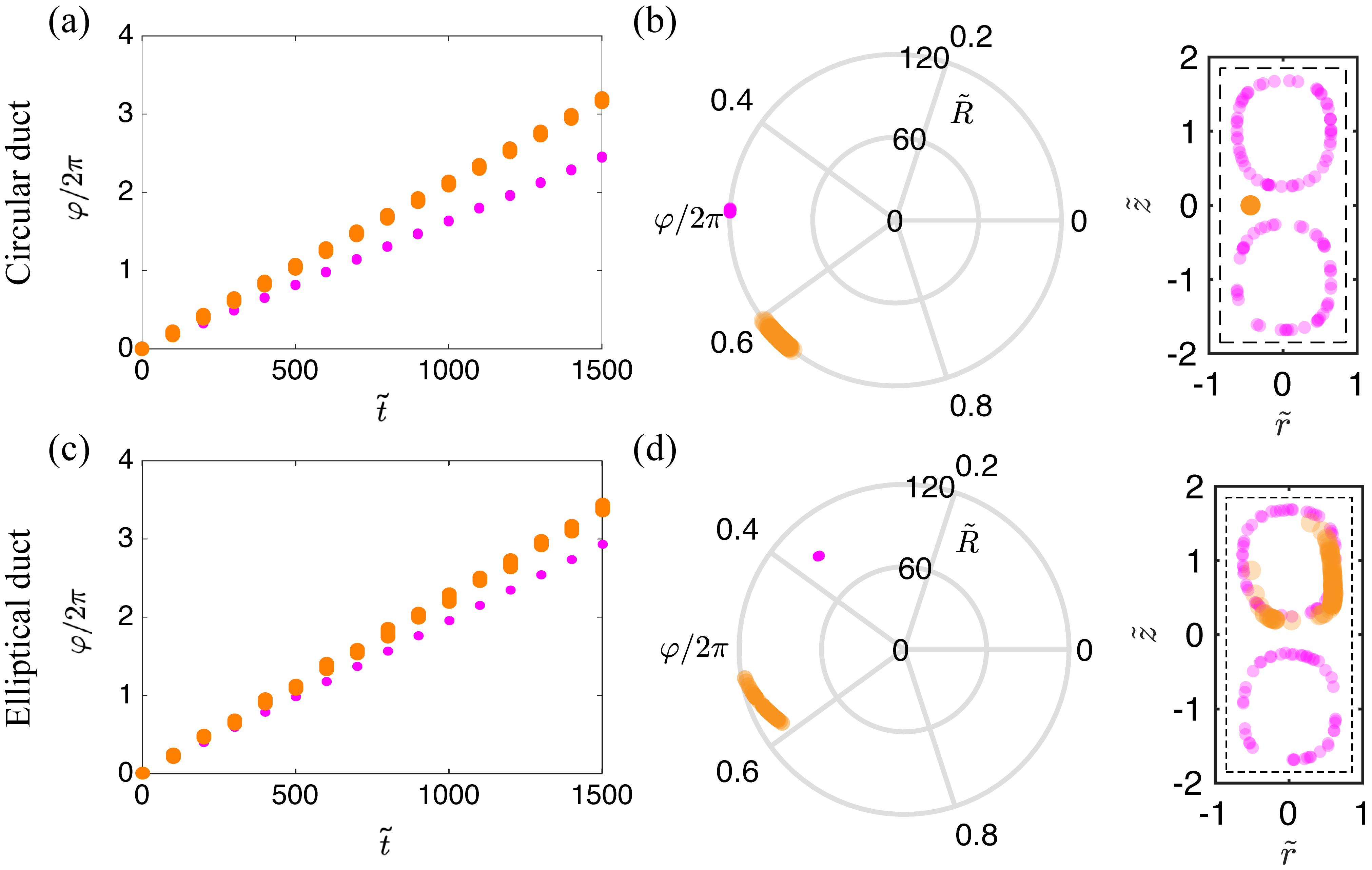}
  \caption{(a,c) Particle position in the longitudinal 
  direction, $\varphi/2\pi$, as a function of time, $\tilde{t}$, for $100$ small particles ($\tilde{a}=0.05$, magenta circles) and $100$ large particles ($\tilde{a}=0.15$, orange circles) for low flow Reynolds number $Re=50$ in (a) circular ($\tilde{a}_e=120$, $e=0$) and (c) elliptical ($\tilde{a}_e=120$, $e=0.75$) ducts. (b,d) The corresponding particle positions in both the polar $(\tilde{R},\varphi)$ and cross-sectional $(\tilde{r},\tilde{z})$ planes at times (b) $\tilde{t}=300$ for the circular duct and (d) $\tilde{t}=500$ for the elliptical duct. The dashed black lines in the cross-sectional plots denote locations of particle centres having surfaces that touch the wall of the duct (solid black lines). The complete simulation of particle separation over the time interval $\tilde{t}\in [0, 1500]$ is given in Supplementary Movies~\ref{SM4} and~\ref{SM5} for the circular and elliptical ducts, respectively.}\label{f7}
\end{figure}

Building on the established mechanisms of longitudinal clustering in an elliptical duct, this section examines how such clustering aids longitudinal particle separation. Particles of different sizes are considered, motivated by their relevance to size-based separation applications such as the isolation of circulating tumour cells from blood \cite{sun2012double,hou2013isolation,warkiani2014slanted}, where tumour cells are typically larger than healthy blood cells. For this purpose, we initialise $100$ large particles ($\tilde{a} = 0.15$) and $100$ small particles ($\tilde{a} = 0.05$) from the same locations in 
the cross-section at $\varphi(0)=0$. We consider two ducts with $\tilde a_e=120$, a circular duct ($e=0$) and an elliptical duct ($e=0.75$), Fig.~\ref{f7}. As mentioned in Section~\ref{PC}, large particles and small particles are distinguished throughout this section by using different marker sizes and colours. 

\begin{figure}[!ht]
  \centering
  \includegraphics[width=0.9\textwidth]{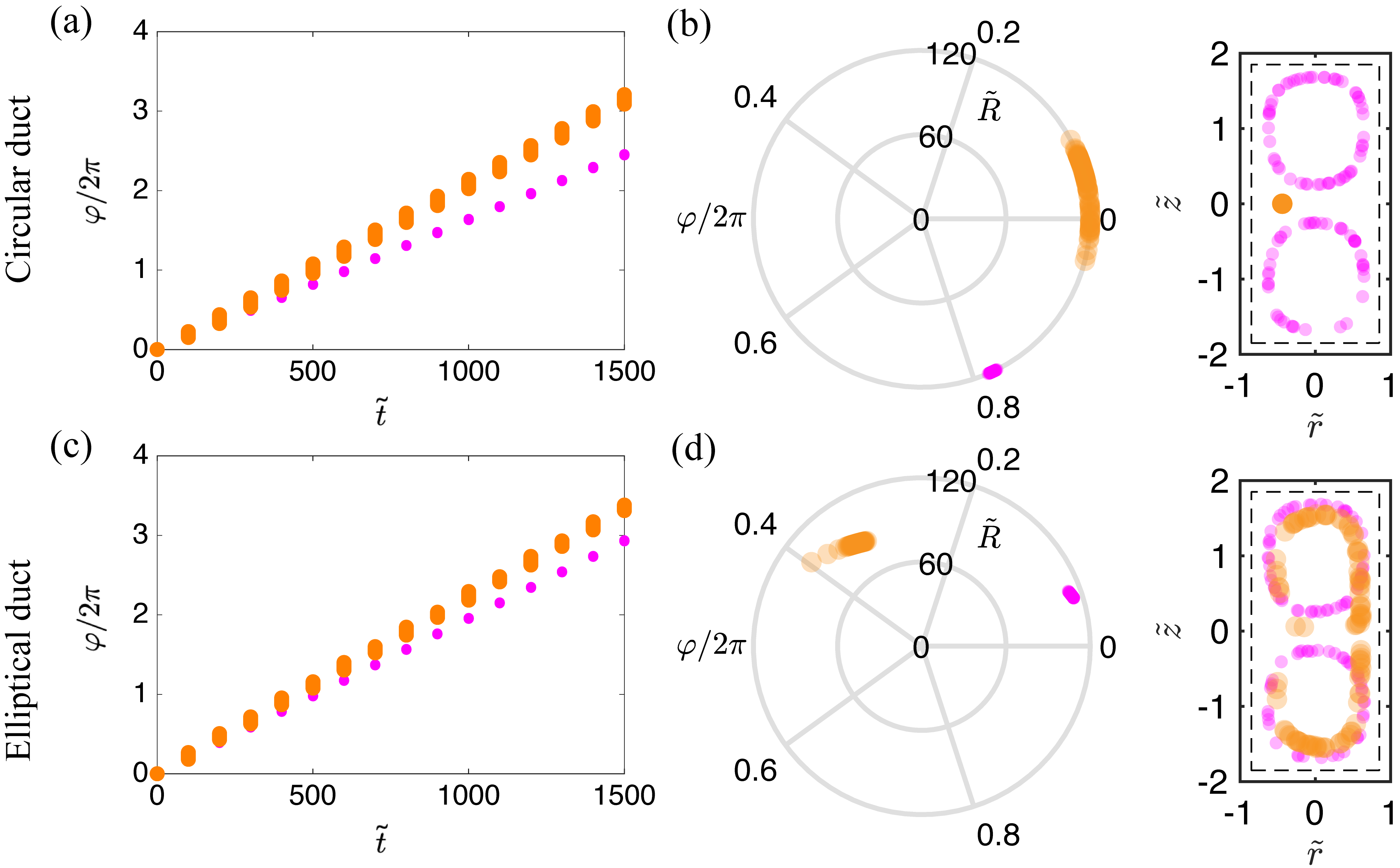}
  \caption{Particle separation dynamics as shown in Fig.~\ref{f7}, but for the larger flow Reynolds number $Re=150$. (a,c) Positions of large ($\tilde a=0.15$, orange circles) and small ($\tilde a=0.05$, magenta circles) in the longitudinal direction, $\varphi/2\pi$, versus time, $\tilde t$, in the (a) circular ($\tilde{a}_e=120$, $e=0$) and (c) elliptical ($\tilde{a}_e=120$, $e=0.75$) ducts. (b,d) The corresponding particle positions in both the polar $(R,\varphi)$ and cross-sectional $(\tilde r,\tilde z)$ planes at times  (b) $\tilde{t}=500$ for the circular duct and (d) $\tilde{t}=1000$ for the elliptical duct. The longitudinal separation is significantly slower in the elliptical duct, but the clustering of the larger particles is tighter.vThe complete simulation of particle separation over the time interval $\tilde{t}\in [0, 1500]$ is given in Supplementary Movies~\ref{SM6} and~\ref{SM7} for the circular and elliptical ducts, respectively.}\label{f8}
\end{figure}

We first examine the effect of duct centreline geometry on longitudinal particle separation at a low Reynolds number of $Re = 50$. 
For the circular duct, Fig.~\ref{f7}(a) shows the particle locations in the longitudinal direction $\varphi/2\pi$, i.e., the number of full rotations along the duct centreline, as a function of $\tilde{t}$. The larger particles travel faster than the smaller particles, resulting in longitudinal separation between the two particle sizes. Moreover, the longitudinal separation increases monotonically with time. On observing the particle separation in both $(\tilde{R},\varphi)$ and $(\tilde{r},\tilde{z})$ planes at $\tilde{t}=300$, Fig.~\ref{f7}(b), we see a separation by size in the cross-section with the larger particles focused near the inner wall, and the smaller particles 
confined in limit cycles, in addition to the longitudinal separation. This cross-sectional separation occurs well before the longitudinal separation emerges (see Supplementary Movie~\ref{SM4}). For the elliptical duct with $e=0.75$, Fig.~\ref{f7}(c) shows similar longitudinal separation to that shown in Fig.~\ref{f7}(a); however, the longitudinal separation is slower as compared to the circular duct. In contrast to the circular duct, there is no cross-sectional separation by size, as seen from Fig.~\ref{f7}(d), due to the larger particles undergoing SNIPER bifurcations; see also Supplementary Movie~\ref{SM5}. We note that for the same major axis radius ($\tilde a_e=120$) but sufficiently small eccentricities ($e<0.35$), the separation behaviour is similar to that for the circular duct.

Fig.~\ref{f8} examines the particle separation dynamics in the same duct as in Fig.~\ref{f7}, but at the larger flow Reynolds number $Re=150$. For the circular duct (Fig.~\ref{f8}(a,b); see also Supplementary Movie~\ref{SM6}), the separation dynamics of the two particle sizes are similar to those shown in Fig.~\ref{f7}(a,b). However, the larger particles become distributed over a longer length of the duct compared with the $Re=50$ case. This arises from pronounced inertial effects, which reduce the rate of particle motion in the cross-section. Once the large particles converge close to the inner wall, the length of the large-particle distribution remains unchanged. For the elliptical duct of Fig.~\ref{f8}(c,d), the particle separation dynamics are also similar to those shown in Fig.~\ref{f7}(c,d). Specifically, cross-sectional separation is not achieved, but longitudinal separation remains possible owing to the clustering of large particles induced by a SNIPER bifurcation (see Supplementary Movie~\ref{SM7}). Note that the longitudinal clustering of the large particles is significantly tighter than for the circular duct, which might be desirable in practice.

\section{Summary}\label{Con}

We have investigated both cross-sectional and longitudinal separation of spherical particles suspended in a three-dimensional curved duct with an elliptical centreline and a rectangular cross-section with a height that is twice the width. The geometry is elliptical except for a small increase in height to prevent self-intersection after a full revolution. 
Unlike other geometries explored to date, the elliptical duct exhibits a periodically varying radius of curvature along its length, which influences both the flow structure and particle motion. The disturbance flow field was computed locally at each angular position using a combination of perturbation and numerical methods, under the assumption that the local radius of curvature varies slowly relative to that of a circular duct with constant bend radius. Further, we evaluated the cross-sectional forces and viscous drag, which are then incorporated into a first-order trajectory model to capture the particle motion within the cross-section and along the duct. 

We examined the influence of duct geometry on particle equilibria and their cross-sectional dynamics. In contrast to a circular duct with constant radius (eccentricity $e=0$), particles travelling through an elliptical duct with $0<e<1$ experience a periodic modulation of equilibrium points, which in turn alters their cross-sectional and longitudinal motion. For 
an elliptical duct with a periodic SNIPER bifurcation, we observe longitudinal clustering that weakens due to the disappearance of a stable node near the inner wall of the cross-section. However, this clustering is subsequently strengthened as particles become vertically localised near the outer duct wall. Quantifying this aggregation using the Kuramoto order parameter reveals strong longitudinal clustering within the SNIPER regime. On reducing the particle size and using a geometrically similar duct featuring a periodic SNIPER bifurcation, we observed a significantly stronger longitudinal clustering as compared to the larger particles. Moreover, this clustering remains unaffected for varying major axis radii and Reynolds numbers. 

Further, we also investigated longitudinal particle separation by size along the elliptical duct, a phenomenon relevant for applications such as isolating tumour cells from healthy blood cells.  To illustrate this, particles of two distinct sizes were suspended in a duct with elliptical centreline, $0\le e < 1$, allowing us to examine the role of centreline geometry across different flow Reynolds numbers. Our analysis demonstrates that simultaneous cross-sectional and longitudinal separation can be achieved for flows at low Reynolds number through a duct with a circular centreline, $e=0$.  This is also true for ducts with small eccentricities, provided the stable node remains close to the inner wall of the duct. However, ducts with larger eccentricities enable longitudinal separation with tighter longitudinal particle clustering, while cross-sectional separation is lost,  due to the periodic occurrence of SNIPER bifurcations. Thus, geometric modulation of the radius of curvature of the duct centreline enables efficient size-based particle separation in the primary longitudinal flow direction only, particularly at small flow Reynolds numbers. At larger Reynolds numbers, these effects still exist, but with a weaker aggregation of the particles in the primary flow direction. 

This study establishes a clear foundation for the design of microfluidic devices that achieve particle separation along the primary flow direction. We have shown that even a simple duct geometry, with small or no periodic variations in curvature, can robustly enable both cross-sectional and longitudinal separation across a wide range of Reynolds numbers; for larger periodic curvature variation, cross-sectional separation is lost while longitudinal separation takes a longer time but is stronger. Thus, longitudinal particle separation may provide an effective strategy for 
particle separation, with strong potential for applications in biomedical and industrial settings.

It may be of peripheral interest to point out that our initial interest in investigating inertial focusing in an elliptical geometry about a SNIPER bifurcation was motivated by the idea that some of the complex behaviour observed in other systems involving coupled oscillators might be observed.
In this case, we posited that the oscillation of forces driving particle motion through each turn of the ellipse is weakly coupled to the oscillatory motion of smaller particles which spend much of their time on limit cycles.
We were particularly interested in whether synchronisation might occur and if such phenomena might be exploited to enhance particle separation.
Alas, we did not observe evidence of synchronisation over the timescales shown within this study.
While it is possible that synchronisation may occur over much larger timescales, such timescales appear to be well beyond what might be practically exploited/utilised.

\vspace{0.5cm}

\begin{center}
\begin{Large}
{\sc Acknowledgments}
\end{Large}
\end{center}
This research is supported under the Australian Research Council’s Discovery Projects funding scheme (project number DP200100834). The results were computed using supercomputing resources provided by the Phoenix HPC service at the University of Adelaide and the R$\bar{\text{a}}$poi HPC service at Victoria University of Wellington.

\vspace{0.5cm}

\begin{center}
\begin{Large}
{\sc Data Availability Statement}
\end{Large}
\end{center}
The data supporting the findings of this study, as well as the MATLAB code and supplementary videos used to generate and illustrate the results, are available from the corresponding author upon request. 


\newpage
\begin{center}
\begin{LARGE}
{\sc Supplementary Information}
\end{LARGE}
\end{center}

\setcounter{figure}{0}
\renewcommand{\thefigure}{S\arabic{figure}}
\setcounter{equation}{0}
\renewcommand{\theequation}{S\arabic{equation}}

\setcounter{section}{0}
\renewcommand{\thesection}{S\arabic{section}}

\section{Modelling a particle suspended in a duct with a constant radius of curvature}\label{App1}
Following \citet{harding2019effect}, we briefly discuss the forces acting on a particle at position $\boldsymbol{x}_p$ suspended in a fluid of viscosity $\mu$ inside a duct of constant curvature, equivalently represented by a circle of bend radius $R_c$. We solve this problem under the assumption that the flow is fully developed, neglecting the effects of any inlets and outlets.

We start with choosing the reference length and velocity scales as $l=\min\{W, H\}= W$ and $U_m$, representing the cross-sectional width and maximum axial velocity of the background flow $\bar{\boldsymbol{u}}$, respectively. We scale the physical parameters as follows: the position $\boldsymbol{x}$ with $a$ (particle radius), velocities with $U_ma/l$, pressure with $\mu U_m/l$, and time with $l/U_m$. We define the interior of the duct as domain $\mathcal{D}$, its boundaries as $\partial\mathcal{D}$, the fluid domain excluding the particle as $\mathcal{F}:=\{\boldsymbol{x} \in \mathcal{D} : |\boldsymbol{x} - \boldsymbol{x}_p| >  1\}$ and the particle surface as $\partial\mathcal{F} \backslash \partial\mathcal{D} := \{\boldsymbol{x} : |\boldsymbol{x} - \boldsymbol{x}_p| = 1\}$. In addition to this, we adopt a reference frame rotating about the vertical $z-$axis with the constant rate $\Theta:=\partial\varphi_p/\partial t$ (scaled with $U_m/l$). The presence of a particle in the duct creates a disturbance velocity $\boldsymbol{v}$ and pressure $q$ field, which is the difference between those fields in the presence and absence of the particle. Thus, the corresponding dimensionless, quasi-steady, Navier-Stokes governing equations and their associated boundary conditions are given as follows:
\begin{subequations}
\begin{align}
-\boldsymbol{\nabla}q+\nabla^2\boldsymbol{v}&=Re_p\big[(\boldsymbol{v}+\bar{\boldsymbol{u}}-\Theta(\boldsymbol{e}_z\times \boldsymbol{x}))\cdot\boldsymbol{\nabla}\boldsymbol{v}+\boldsymbol{v}\cdot \boldsymbol{\nabla}\bar{\boldsymbol{u}}+\Theta (\boldsymbol{e}_z\times \boldsymbol{v})\big]~~\mbox{on}~~\boldsymbol{x}\in \mathcal{F}, \label{A1a}\\
\boldsymbol{\nabla}\cdot\boldsymbol{v}&=0~~\mbox{on}~~\boldsymbol{x}\in \mathcal{F},\label{A1b}\\
\boldsymbol{v}&=0~~\mbox{on}~~\boldsymbol{x}\in \partial\mathcal{D},\label{A1c}\\
\boldsymbol{v}&=-\bar{\boldsymbol{u}}+\boldsymbol{\Omega}_p\times (\boldsymbol{x}-\boldsymbol{x}_p)~~\mbox{on}~~\boldsymbol{x}\in \partial \mathcal{F}\backslash\partial D .\label{A1d}
\end{align}\label{A1}
\end{subequations}
\hspace{-0.1cm}Here, $\Omega_p$ denotes the particle's spin (scaled with $U_m/l$); $\boldsymbol{e}_z$ denotes the unit vector in the $z-$direction; the particle Reynolds number is given as $Re_p=a^2Re/l^2$, with $Re=\rho U_ml/\mu$ being the flow Reynolds number corresponding to the fluid of density $\rho$. 

The particle experiences the force $\boldsymbol{F}$ and torque $\boldsymbol{T}$ arising from the disturbed flow field, supplemented by the corresponding centripetal force and gyroscopic torque. The resulting expressions are given as follows:
\begin{align}
\boldsymbol{F}&=-\frac{4\pi}{3}\Theta^2 (\boldsymbol{e}_z\times(\boldsymbol{e}_z\times\boldsymbol{x}_p))+\int_{{|\boldsymbol{x}-\boldsymbol{x}_p|<1}}\bar{\boldsymbol{u}}\cdot \boldsymbol{\nabla}\bar{\boldsymbol{u}}~dV\nonumber\\
&+\frac{1}{Re_p}\int_{|\boldsymbol{x}-\boldsymbol{x}_p|=1} -\boldsymbol{n} \cdot (-q\boldsymbol{I}+\boldsymbol{\nabla}\boldsymbol{v}+\boldsymbol{\nabla}\boldsymbol{v}^\mathrm{T}) dS,\label{A2}\\
\boldsymbol{T}&=-\frac{8\pi}{15}\Theta (\boldsymbol{e}_z\times\Omega_p)+\int_{{|\boldsymbol{x}-\boldsymbol{x}_p|<1}}(\boldsymbol{x}-\boldsymbol{x}_p)\times (\bar{\boldsymbol{u}}\cdot \boldsymbol{\nabla}\bar{\boldsymbol{u}})~dV\nonumber\\
&+\frac{1}{Re_p}\int_{|\boldsymbol{x}-\boldsymbol{x}_p|=1} (\boldsymbol{x}-\boldsymbol{x}_p)\times (-\boldsymbol{n} \cdot (-q\boldsymbol{I}+\boldsymbol{\nabla}\boldsymbol{v}+\boldsymbol{\nabla}\boldsymbol{v}^\mathrm{T})) dS,\label{A3}
\end{align}
where $-\boldsymbol{n}$ is the normal vector pointing outward from the particle centre. Now, we perform perturbation expansions on the disturbance fields by assuming $Re_p$ to be small, specifically
\begin{subequations}
    \begin{align}
        \boldsymbol{v}=\boldsymbol{v}_0+Re_p \boldsymbol{v}_1+\mathcal{O}(Re_p^2),\\
        \boldsymbol{q}=\boldsymbol{q}_0+Re_p \boldsymbol{q}_1+\mathcal{O}(Re_p^2),
    \end{align}\label{A4}
\end{subequations}
\hspace{-0.1cm}which are substituted in Eqs.~\eqref{A1}, resulting in the leading-order equations for $\boldsymbol{v}_0$ and $q_0$ and the first-order equations for $\boldsymbol{v}_1$ and $q_1$. The leading-order equations correspond to the Stokes equations subject to the non-zero boundary conditions, while the first-order equations capture the inertial effects on the particle. On substituting the above Eqs.~\eqref{A4} in the expression for force \eqref{A2}, we get the following cross-sectional  forces to leading-order:
\begin{align}
       F_r=\boldsymbol{e}_r\cdot (Re_p^{-1}\boldsymbol{F}_{-1,s}+\boldsymbol{F}_0)~~~~~\mbox{and}~~~~       F_z=\boldsymbol{e}_z\cdot (Re_p^{-1}\boldsymbol{F}_{-1,s}+\boldsymbol{F}_0), \label{A5}
\end{align}
where $\boldsymbol{F}_{-1,s}$ corresponds to the drag arising from the background flow, whereas $\boldsymbol{F}_0$ denotes the inertial lift force. The cross-sectional forces ($F_r$ and $F_z$) and their corresponding drags ($C_r$ and $C_z$) are obtained from a finite element method, which is detailed in \citet{harding2019effect}. 

\section{Axial velocity component of the first-order model}\label{App2}

In this section, we derive the axial component of the first-order model as given in Eq.~\eqref{e4}. Consider a particle advected along the duct centreline from one axial location to another. A small change in the orbital coordinate $\varphi{_p}$ produces the displacement
\begin{align}
     \frac{d\boldsymbol{x}_p}{d\varphi_p} =R'(\varphi_p) \boldsymbol{e}_r|_{\varphi=\varphi_p}+R(\varphi_p) \frac{d\boldsymbol{e}_r}{d\varphi}\bigg|_{\varphi=\varphi_p},\label{A6}
\end{align}
where $\boldsymbol{e}_r=\cos{\varphi}~\boldsymbol{e}_x+\sin{\varphi}~\boldsymbol{e}_y$, the prime denotes $d/d\varphi$, and $R(\varphi)$ is the radius of the duct centreline, Eq.~\eqref{e2}. The associated arclength element is obtained by taking the norm of \eqref{A6},
\begin{align}
ds|_{\varphi=\varphi_p}=\bigg\lVert\frac{d\boldsymbol{x}_p}{d\varphi_p}\bigg\rVert d\varphi_p=\sqrt{(R'(\varphi_p))^2+(R(\varphi_p))^2}~d\varphi_p,\label{A7}
\end{align}
where $\sqrt{(R'(\varphi_p))^2+(R(\varphi_p))^2}$ is the metric factor converting the angular change, $d\varphi$, into the corresponding arclength change, $ds$. Dividing Eq.~\eqref{A7} by $dt$ and rearranging yields the evolution of the orbital parameter:
\begin{align}
    \frac{d\varphi_p}{dt}=\frac{\bar{u}_{\varphi_p}}{\sqrt{(R'(\varphi_p))^2+(R(\varphi_p))^2}}, \label{A8}
\end{align}
where $\bar{u}_{\varphi_p}=ds/dt|_{\varphi=\varphi_p}$ is the axial component of the background flow.

\begin{figure}[b!]
  \centering
  \includegraphics[width=0.9\textwidth]{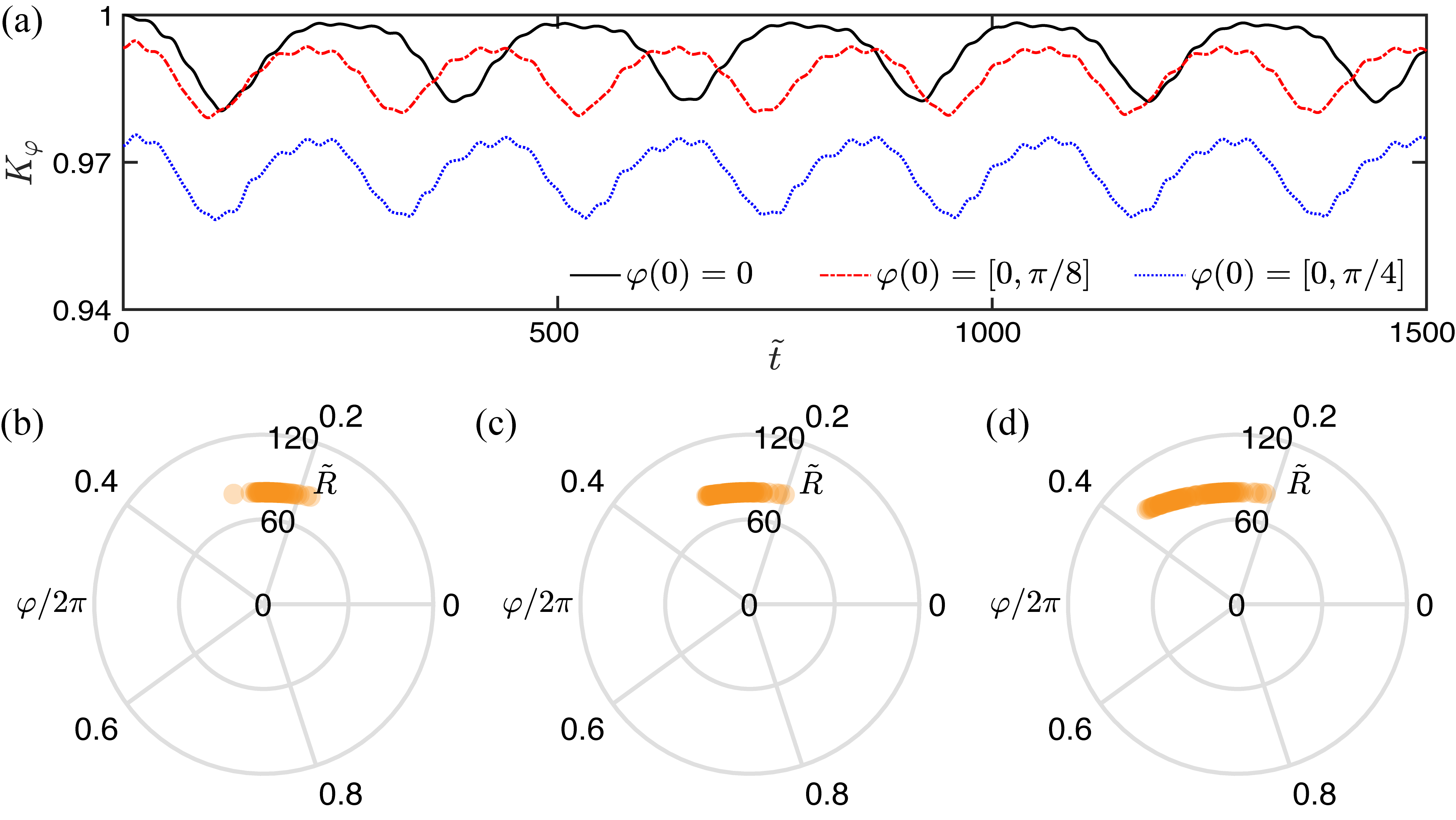}
  \caption{(a) Kuramoto order parameter $K_{\varphi}$ (averaged over $N=100$ particles) as a function of $\tilde{t}$, showing the effect of an initial uniform orbital particle distribution over $\varphi(0)\in[0,\varphi_\ell]$, for given $\varphi_\ell$, within a flow of $Re=150$ in an elliptical duct with $\tilde{a}_e=120$ and $e=0.75$. 
  (b--d) The particle distribution in the polar $(R,\varphi)$ plane at $\tilde{t}=250$ for (b) $\varphi(0)=0$, (c) $\varphi(0)=[0,\pi/8]$ and (d) $\varphi(0)=[0,\pi/4]$.}\label{f6_n}
\end{figure}

\section{Particle clustering for different axial initial condition}\label{App4}

In practice, particles will be released into the elliptical duct along a finite region $\varphi\in [0,\varphi_{\ell}]$, with an initial spatial distribution both in the cross-section and in the longitudinal ($\varphi$) direction. Here, we investigate how a uniform longitudinal particle distribution 
affects longitudinal particle clustering as quantified by the Kuramoto order parameter $K_\varphi$, within a flow of Reynolds number $Re=150$ in an elliptical duct with $\tilde a_e=120$, $e=0.75$. Note that the value of $K_{\varphi}$ at time $\tilde{t}=0$ depends only on the prescribed initial distribution of particles in the longitudinal direction. Fig.~\ref{f6_n} compares the release at $t=0$ of 100 particles within a single longitudinal cross-section $\varphi(0)=0$, with the release of $100$ particles which, at $t=0$, are uniformly distributed over the intervals $\varphi(0)\in[0,~\pi/8]$ and $\varphi(0)\in[0,~\pi/4]$. As $\tilde{t}$ increases, the elliptical duct exhibits behaviour similar to that in Fig.~\ref{f5}, with periodic and strong particle aggregation in the longitudinal direction. However, increasing the initial particle distribution in the flow direction leads to a noticeable weakening of longitudinal clustering. This is obvious from the particle positions in the $(\tilde{R},\varphi)$ plane at $\tilde{t}=250$ (Fig.~\ref{f6_n}(b–d)), where $K_{\varphi}(\tilde{t})$ attains its maximum value. Furthermore, the periodic maxima of $K_{\varphi}(\tilde{t})$ for $\tilde{t}>0$ coincide with their respective initial values $K_{\varphi}(0)$ as shown in Fig.~\ref{f6_n}(a).

\section{Bifurcation diagram and clustering of small particles}\label{App3}

\begin{figure}[t!]
  \centering
  \includegraphics[width=0.9\textwidth]{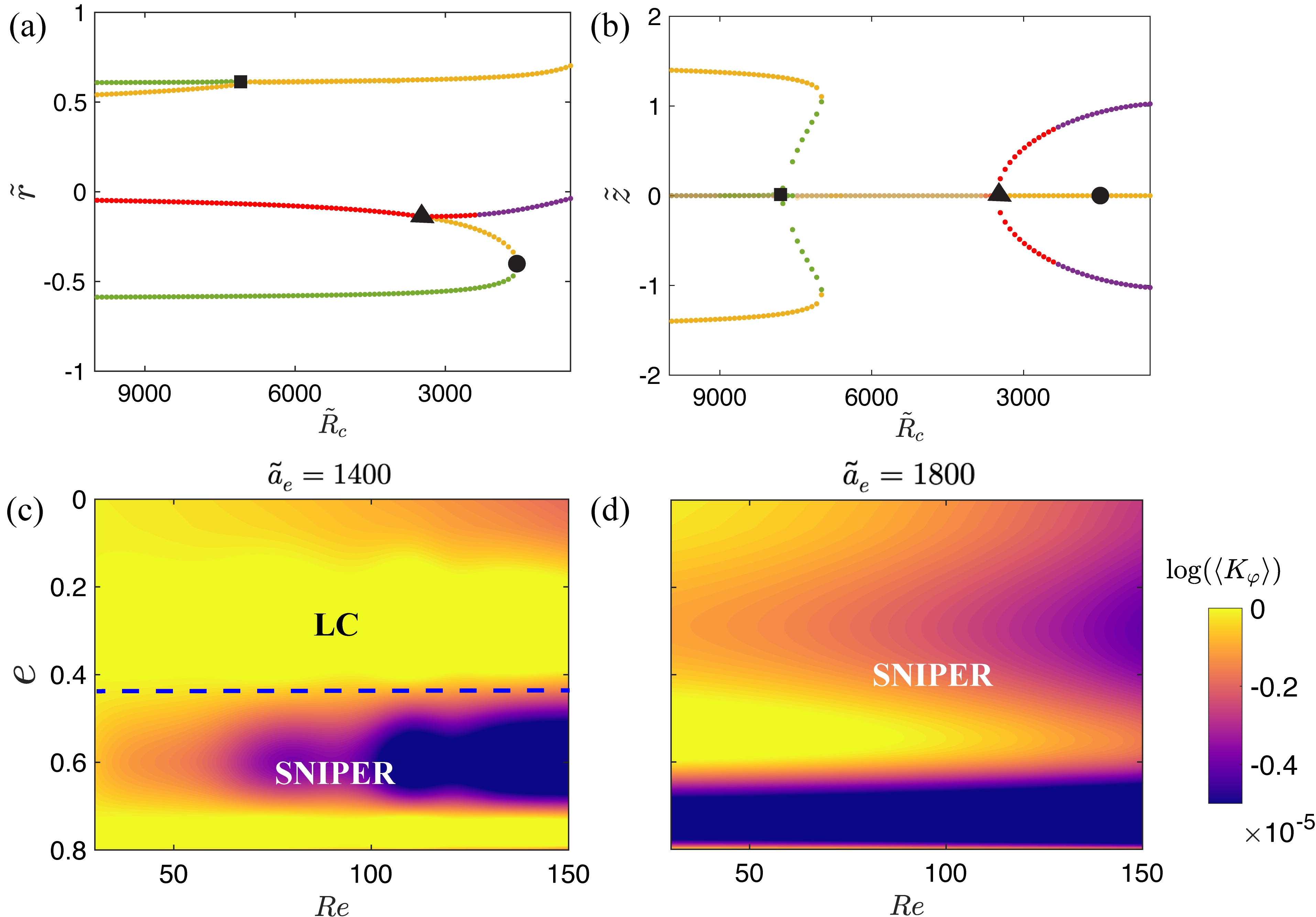}
  \caption{(a,b) Bifurcation diagrams showing the (a) horizontal $\tilde r$ and (b) vertical $\tilde z$ coordinates of equilibria as a function of the radius of curvature $\tilde R_c$, which decreases from left to right.  The particle is of small radius $\tilde{a}=0.05$ and the duct has a $1\times2$ rectangular cross-section. The stable nodes, saddle points, unstable nodes, and unstable spiral points are shown in green, yellow, red, and purple, respectively. The black squares, triangles, and circles denote the locations of subcritical pitchfork, supercritical pitchfork, and saddle-node infinite-period (SNIPER) bifurcations, respectively. (c,d) Contour maps showing the log of time-averaged Kuramoto order parameter, $\log(\langle K_{\varphi}\rangle)$ over $\tilde{t}=1500$, as a function of eccentricity, $e$, and flow Reynolds number, $Re$, for major axis radii (c) $\tilde{a}_e=1400$, and (d) $\tilde{a}_e=1800$. LC denotes the region in which there are stable limit cycles; SNIPER denotes the region in which the stable node and saddle point combination undergoes infinite-period bifurcations; the blue-dashed line marks the division between these two regions.} \label{sf1}
\end{figure}

In this section, we investigate the bifurcation diagram and the time-averaged Kuramoto parameter corresponding to a small particle with $\tilde{a}=0.05$. In Fig.~\ref{sf1}(a,b), we examine the bifurcation diagrams showing the $\tilde r$ and $\tilde z$ coordinates of the equilibria versus $\tilde R_c$, for a circular duct with constant radius of curvature ($\tilde{R}_c=\tilde{a}_e$ and $e=0$). The representation and nature of stable nodes, saddle points, unstable nodes, and unstable spirals are the same as in Fig.~\ref{f2}(a,b) for the large particle, $a_c=0.15$. As for the large particle, the small particle exhibits the same sequence of bifurcation events with decreasing radius of curvature, $\tilde{R}_c$: a subcritical pitchfork bifurcation (black square), followed by a supercritical pitchfork bifurcation (black triangle), and ultimately a saddle-node infinite-period (SNIPER) bifurcation (black circle). Notably, for the smaller particle radius, $\tilde{a}=0.05$, this entire sequence is shifted toward significantly larger radii of curvature, occurring over the range $\tilde{R}_c \in [10^3,10^4]$.

For this small particle, we also investigate the log of the time-averaged Kuramoto order parameter, $\log(\langle K_{\varphi}\rangle)$, in the $(e, Re)$ plane for varying major axis radii $\tilde{a}_e$, Fig.~\ref{sf1}(c, d). Unlike the case for the large particle ($\tilde a=0.15$) of Fig.~\ref{f4}(c, d), the value of $\langle K_{\varphi} \rangle$ is  
close to unity ($\log(\langle K_{\varphi} \rangle) < 10^{-5}$) 
over the $(e, Re)$ space shown for various $\tilde{a}_e$, implying small particles aggregate strongly in the streamwise direction irrespective of flow and geometric factors. 

\bibliography{REF}

\newpage

\setcounter{figure}{0}
\renewcommand{\figurename}{Supplementary Movie}
\renewcommand{\thefigure}{\arabic{figure}}

\begin{center}
\section*{Supplementary Movies}\label{appG}
\end{center}

\begin{figure}[!h]
\centering
\caption{The file ``Movie 1.mp4'' shows the time-dependent motion of $100$ large particles ($\tilde{a}=0.15$), shown in both $(\tilde{R},\varphi)$ and $(\tilde{r},\tilde{z})$ planes, corresponding to Fig.~\ref{f4}(a,b), over the total duration of $\tilde{t}=1500$ for the circular duct.}
\label{SM1}
\end{figure}

\begin{figure}[!h]
\centering
\caption{The file ``Movie 2.mp4'' shows the time-dependent motion of $100$ large particles ($\tilde{a}=0.15$) shown in both $(\tilde{R},\varphi)$ and $(\tilde{r},\tilde{z})$ planes, corresponding to Fig.~\ref{f4}(c,d), over the total duration of $\tilde{t}=1500$ for the elliptical duct.}
\label{SM2}
\end{figure}

\begin{figure}[!h]
\centering
\caption{The file ``Movie 3.mp4'' shows the time-dependent motion of $100$ small particles ($\tilde{a}=0.05$), shown in both $(\tilde{R},\varphi)$ and $(\tilde{r},\tilde{z})$ planes, corresponding to Fig.~\ref{f6}(a,b), over the total duration of $\tilde{t}=1500$ for the elliptical duct. }
\label{SM3}
\end{figure}

\begin{figure}[!h]
\centering
\caption{The file ``Movie 4.mp4'' shows the time-dependent motion of $100$ small particles ($\tilde{a}=0.05$, magenta circles) and $100$ large particles ($\tilde{a}=0.15$, orange circles), shown in both $(\tilde{R},\varphi)$ and $(\tilde{r},\tilde{z})$ planes, corresponding to Fig.~\ref{f7}(a,b), over the total duration of $\tilde{t}=1500$ for the circular duct with low Reynolds number, $Re=50$.}
\label{SM4}
\end{figure}

\begin{figure}[!h]
\centering
\caption{The file ``Movie 5.mp4'' shows the time-dependent motion of $100$ small particles ($\tilde{a}=0.05$, magenta circles) and $100$ large particles ($\tilde{a}=0.15$, orange circles), shown in both $(\tilde{R},\varphi)$ and $(\tilde{r},\tilde{z})$ planes, corresponding to Fig.~\ref{f7}(c,d), over the total duration of $\tilde{t}=1500$ for the elliptical duct for the circular duct with low Reynolds number, $Re=50$.}
\label{SM5}
\end{figure}

\begin{figure}[!h]
\centering
\caption{The file ``Movie 6.mp4'' shows the time-dependent motion of $100$ small particles ($\tilde{a}=0.05$, magenta circles) and $100$ large particles ($\tilde{a}=0.15$, orange circles), shown in both $(\tilde{R},\varphi)$ and $(\tilde{r},\tilde{z})$ planes, corresponding to Fig.~\ref{f8}(a,b), over the total duration of $\tilde{t}=15000$ for the circular duct for the circular duct with high Reynolds number, $Re=150$.}
\label{SM6}
\end{figure}

\begin{figure}[!h]
\centering
\caption{The file ``Movie 7.mp4'' shows the time-dependent motion of $100$ small particles ($\tilde{a}=0.05$, magenta circles) and $100$ large particles ($\tilde{a}=0.15$, orange circles), shown in both $(\tilde{R},\varphi)$ and $(\tilde{r},\tilde{z})$ planes, corresponding to Fig.~\ref{f8}(c,d), over the total duration of $\tilde{t}=1500$ for the circular duct for the circular duct with high Reynolds number, $Re=150$.}
\label{SM7}
\end{figure}

\end{document}